\documentclass[aps,prb,twocolumn,showpacs,groupedaddress]{revtex4}

\def\comment#1{\bigskip\hrule\smallskip#1\smallskip\hrule\bigskip}   
\usepackage{amsmath}
\usepackage{graphicx}
\usepackage{dcolumn}
\usepackage{bm}

\begin{document}

\title{Quantum Monte Carlo simulation of spin-polarized tritium}

\author{I. Be\v{s}li{\'c}$^1$, L. Vranje\v{s} Marki{\'c}$^{1,2}$, J.
Boronat$^3$}
\affiliation{
$^1$ Faculty of Science, University of Split, HR-21000 Split,
Croatia}
 \affiliation{$^2$ Institut f\"ur Theoretische Physik, 
Johannes Kepler Universit\"at, A 4040 Linz, Austria}
\affiliation{$^3$ Departament de F\'\i sica i Enginyeria Nuclear, Campus Nord
B4-B5, Universitat Polit\`ecnica de Catalunya, E-08034 Barcelona, Spain}
\def\comment#1{\bigskip\hrule\smallskip#1\smallskip\hrule\bigskip}   

\date{\today}

\begin{abstract}
The ground-state properties of spin-polarized tritium T$\downarrow$ at zero temperature
 are
obtained  by means of diffusion Monte Carlo calculations. Using an accurate
{\em ab
initio} T$\downarrow$-T$\downarrow$ interatomic potential we have studied
its liquid phase, from the spinodal point until densities above 
its freezing point.  
The equilibrium density of the liquid is significantly higher and the
equilibrium energy of $-3.664(6)$ K significantly lower than in previous
approximate descriptions. The solid phase has also been studied for three
lattices up to high pressures, and we find that hcp lattice is slightly
preferred. The liquid-solid phase transition has been determined using
the double-tangent Maxwell construction; at zero temperature,
bulk tritium freezes at a pressure of $P=9(1)$ bar.

\end{abstract}

\pacs{67.63.Gh,67.80.fh}

\maketitle

\section{Introduction}

Spin-polarized hydrogen (H$\downarrow$), and its isotopes spin-polarized
deuterium (D$\downarrow$) and tritium (T$\downarrow$), 
are fully quantum systems with a wealth of interesting properties.
The low mass of H$\downarrow$ and the very weak attractive part of
its interaction potential makes possible that it remains in a stable gas
phase even in the
limit of zero temperature. Recent microscopic calculations at zero
temperature have shown that the gas phase persists for pressures up to 170 bar.~\cite{hydrogen}
 The gaseous state of bulk H$\downarrow$ was proposed in
1976 by Stwalley and Nosanow~\cite{StwaleyNosanow}  as the best reliable
option to achieve a Bose-Einstein condensate (BEC) state without the strong
depletion of the condensate that correlations induce in superfluid $^4$He. 
  After
22 years of intense and continued experimental work by many groups, Fried \textit{et
al.}~\cite{fried} managed to achieve a BEC of
H$\downarrow$. Problems that needed to be overcome included recombination
on the walls, by working with a wall-free confinement, and low evaporation
rates by using spin resonance. In the meantime, other BEC states were first
 achieved in 1995 working with  alkali gases such as Rb, Na and Li.~\cite{becalkali}.  

In contrast to alkali gases, the hydrogen-hydrogen interatomic
interaction is very well known and leads to a stable gas phase.  
Spin-polarized hydrogen atoms interact via
the triplet potential $b$ $^3\Sigma_u^+$ determined in an essentially exact
way by Kolos and Wolniewicz,~\cite{kolos} and recently extended to larger
interparticle distances by Jamieson \textit{et al.}.~\cite{jamieson}
The s-wave scattering length $a$ is appreciably smaller than the
typical values for alkalis, which retards evaporative cooling and produces
a higher transition temperature  (50 $\mu K$).

Essentially the same interaction can be applied for heavier isotopes
spin-polarized deuterium D$\downarrow$ and tritium T$\downarrow$. 
D$\downarrow$ atoms obey Fermi statistics, and so the zero-pressure state
of  bulk D$\downarrow$ depends on the number of occupied nuclear spin
states. In the limit of zero pressure and zero temperature, previous
theoretical studies~\cite{panoff,flynn,skjetne} have shown that 
(D$\downarrow_1$) with only one occupied nuclear spin state is a gas, while
bulk D$\downarrow$ with two (D$\downarrow_2$)  and three 
(D$\downarrow_3$)  equally occupied nuclear spin states  remains liquid. 
Spin-polarized tritium, which obeys Bose statistics, is expected to be 
liquid~\cite{etters,miller,tritium} due to its larger mass. In fact, Stwaley and
Nosanow~\cite{StwaleyNosanow} suggested it should behave very much like
liquid $^4$He and therefore constitute a second example of bosonic
superfluid. Recently, microscopic properties of T$\downarrow$ clusters
have been studied by Blume \textit{et al.}~\cite{blume} using the diffusion
Monte Carlo (DMC) method. In addition, in Ref. \onlinecite{blume} spin-polarized
tritium is suggested  as a new BEC gas in optical dipol trap. It has the same advantage of a nearly exact
knowledge of the interatomic potential as spin-polarized hydrogen but,
unlike H$\downarrow$, it has a very broad Feshbach resonance that can be
used to control the condensate in a trap.

In the present work, we present a DMC study of the liquid and solid phases
of spin-polarized tritium. For bosonic many-body systems at zero
temperature DMC methods lead to exact estimations of the ground-state energy 
and related properties within statistical errors. Using the 
\textit{ab initio} T$\downarrow$-T$\downarrow$ interatomic potential within
the DMC method, we report accurate microscopic results for energetic and
structural properties of the bulk system. Relevant results of this work
include the determination of the equilibrium density and energy per particle, the
spinodal point and the liquid-solid phase transition. 

In Sec. II, we briefly describe the DMC method and the trial wave functions
used for importance sampling in the liquid and solid phases. In Sec.
III, the results of the DMC simulations are reported in several
subsections. The first and second one are devoted to the microscopic
results for the liquid and solid phases, respectively. In the last one, we
study the liquid-solid phase transition point and report results on the
freezing and melting densities. Finally, Sec. IV comprises a summary of our results
and the main conclusions.

\section{Method}

The starting point of the DMC method is the Schr\"odinger equation 
written in imaginary time,
\begin{equation}
-\hbar \frac{\partial \Psi(\bm{R},t)}{\partial t} = (H- E_{\text r}) \Psi(\bm{R},t) \ ,
\label{srodin}
\end{equation}
with an $N$-particle Hamiltonian
\begin{equation}
 H = -\frac{\hbar^2}{2m} \sum_{i=1}^{N} \bm{\nabla}_i^2 + \sum_{i<j}^{N} V(r_{ij})  \ .
\label{hamilto}
\end{equation}

In Eq. (\ref{srodin}), $E_{\text r}$ is a constant acting as a reference energy and
$\bm{R} \equiv (\bm{r}_1,\ldots,\bm{r}_N)$ is a \textit{walker} in Monte
Carlo terminology. In order to reduce the variance to a manageable level 
it is a common practice to use importance sampling by introducing a 
trial wave function $\psi(\bm{R})$. Then, the Schr\"odinger equation is
rewritten for the wave function $\Phi(\bm{R},t)=\Psi(\bm{R},t)
\psi(\bm{R})$ and solved in a stochastic form.
 In the limit $t \rightarrow \infty$ only the lowest energy
eigenfunction, not orthogonal to $\psi(\bm{R})$, survives and then the
sampling of the ground state is effectively achieved. Apart from
statistical uncertainties, the energy of a $N$-body bosonic system is
exactly calculated.

The interaction between T$\downarrow$ atoms  is described with the
spin-independent central triplet pair potential $b$$^3\Sigma_u^+$.
It has been determined in an essentially 
exact way by Kolos and Wolniewicz,~\cite{kolos} and recently extended to
larger interparticle distances by Jamieson \textit{et al.}
(JDW).~\cite{jamieson} 
As in a recent DMC calculation of bulk H$\downarrow$,\cite{hydrogen}  
we have used a cubic spline to interpolate between JDW data. This
interaction is then smoothly connected to the long-range behavior of the
T$\downarrow$-T$\downarrow$  potential as
calculated by Yan \textit{et al.}.~\cite{yan}
The JDW potential used in the present work has a core diameter
$\sigma=3.67$~\AA\, and a minimum of $-6.49$ K at a distance $4.14$ \AA.
A comparison  between different potentials employed in the past is 
reported in Ref. \onlinecite{hydrogen}.   We have also verified that the addition of mass-dependent adiabatic corrections (as calculated by Kolos and Rychlewski~\cite{kolos2}) to the JDW potential does not change the energy of the bulk spin-polarized tritium.

The trial wave function used for the simulation of the liquid phase is of
Jastrow form, 
\begin{equation}
 \psi_{\text J}(\bm{R}) = \prod_{i<j}^{N} f(r_{ij}) \ .
 \label{jastrow} 
\end{equation}
The two-body correlation function $f(r)$ is the same as in our
previous study of spin-polarized hydrogen,~\cite{hydrogen}
\begin{equation}
f(r)=\exp [-b_1 \exp(-b_2r )] \ ,
\label{trial}
\end{equation}
where $b_1$ and $b_2$ are variational parameters. The same form was also
used in the variational Monte Carlo (VMC) calculation  of Etters 
\textit{et al.},~\cite{etters} who
modeled the H$\downarrow$-H$\downarrow$ interaction with a Morse potential
fitted to reproduce Kolos and Wolniewicz \textit{ab initio}
data.~\cite{kolos} This analytic form (\ref{trial}) corresponds to the WKB 
solution of the two-body
Schr\"odinger equation for small interparticle distances when the potential
is of Morse type.    

Simulations of the crystalline
bcc, fcc and hcp phases have been also carried out; in this case, we use a Nosanow-Jastrow 
model
\begin{equation}
\psi_{\text{NJ}}(\bm{R}) = \psi_{\text J}(\bm{R}) 
\prod_{i}^{N}h(r_{iI}) \ ,
\label{nosanow}
\end{equation}
 where $h(r)=\exp(-\alpha r^2/2)$ is a localizing function which links every
particle $i$ to a fixed lattice point $\bm{r}_I$. The parameter $\alpha$ is
optimized variationally.

The trial wave function $\psi(\bm{R}) $ has been optimized  for 
the density range where the equation of state has been calculated, by using the VMC method. 
The liquid phase has been studied for densities in the
interval from $0.006$ \AA$^{-3}$  to $0.02$ \AA$^{-3}$.  Within this
interval, the value of the parameter
$b_1$ (\ref{trial}) that optimizes the trial wave function takes 
increasing values with the density from 110 to 180,
while the second parameter $b_2$ (\ref{trial}) does not change significantly, assuming values
from $1.28$   to $1.35$ \AA$^{-1}$ . For the three solid lattices (bcc, fcc
and hcp), the calculations have been carried out 
from $0.008$  to $0.024$ \AA$^{-3}$. As in the case of the liquid phase,
the parameter $b_2$ slightly changes, from $1.29$  to $1.45$ \AA$^{-1}$.  
The parameter $b_1$ increases with density from the melting point up to the
highest density studied here, taking values  
from 80 to 148.  The parameter $\alpha$, which models the strength of the
localization of particles around the lattice sites, increases with density.
In the case of the bcc phase, optimized values  of $\alpha$ range from
$0.33$   to
$1.80$ \AA$^{-2}$, for the fcc phase from $0.32$   to $2.47$ \AA$^{-2}$, and
for the hcp phase from $0.28$   to $2.21$ \AA$^{-2}$.
The statistical errors of the 
variational energies in this optimization procedure with VMC are compatible with those of the 
DMC results (see Tables \ref{tab:Liquid} and \ref{tab:HCPtritium}).
  
We use the DMC method accurate to second order in the time step $\Delta
t$,~\cite{boro} which allows us to use larger $\Delta t$ values than in
linear DMC. Both  the time-step dependence and the mean walker population
have been studied carefully in order to eliminate any bias coming from
them. 

Any simulation of a bulk system with a finite number of particles requires
a size-dependence analysis in order to achieve results as close as possible to
the thermodynamic limit. For
the liquid phase, we have used 250 particles in all simulations and checked
at the VMC level that with the
addition of standard tail corrections, the size dependence of the energy
remains smaller than the typical size of the statistical error. 
On the other hand, in all the
solid state simulations we have assumed periodic boundary conditions, with
256, 250, and 180 atoms for the bcc, fcc, and hcp lattices, respectively. 
Due to the periodic order of the solid,  standard tail
corrections which assume ($g(r)=1$) beyond $r > L/2$, where $L$ is the length
of the simulation box, become rather inaccurate. 
In order to better determine the energy tail corrections we
have studied the size dependence of the energy at the VMC level, where
larger number of particles can be used. From the VMC results one extracts
the tail corrections for a given number of atoms and then these are added
to the DMC energies.  With this procedure it was possible to  reproduce
accurately the experimental equation of state of solid
$^4$He~\cite{overpressure}.

\section{Results}

\subsection{Liquid phase}

Spin-polarized T in its liquid phase has been studied in the density
range from spinodal point up to densities above crystallization. DMC
results for the total and kinetic energy per  particle at different
densities are reported in Table \ref{tab:Liquid}. In order to remove any
residual bias from the trial wave function, kinetic energies are calculated
as differences between  total energies and pure estimations of potential
energies. The total energy is negative approximately up to the density
$\rho=$ 0.012 \AA$^{-3}$. The potential energy per particle is negative in
all the density regime studied, presenting a minimum value of around -19 K at the
density $\rho=$ 0.014 \AA$^{-3}$

\begin{table}
\begin{center}
\begin{ruledtabular} 
\begin{tabular}{lrrrr}         $\rho$ (\AA $^{-3}$) & 
~$E/N$ (K)    & ~~$T/N$ (K)    & ~~$P$ (bar)   & $c$ (m/s)     \\ \hline 
0.006              &   -3.427(2)        &     5.578(17)   &   -1.43(2)     &    71(3)  \\
0.0074             &   -3.664(6)        &     7.778(18)   &   -0.12(1)     &   189(3)  \\
0.009              &   -3.320(7)        &     10.84(3)    &    5.11(6)     &   318(3)         \\         
0.01               &   -2.675(8)        &     13.06(4)    &   11.6(1)      &   402(4)  \\
0.0125             &    0.86(2)         &     19.28(6)    &   45.2(4)      &   630(5)        \\
0.016              &    12.26(6)        &     30.33(10)    &   161(2)       &   992(8)  
\end{tabular} 
\end{ruledtabular}
\end{center}
\caption{Results for liquid
T$\downarrow$ at different densities $\rho$: energy per particle ($E/N$), 
kinetic energy per particle ($T/N$), pressure ($P$), and speed of sound 
($c$). Figures in parenthesis are the statistical errors.} 
\label{tab:Liquid}
\end{table}

\begin{figure}
\centering
        \includegraphics[width=0.8\linewidth]{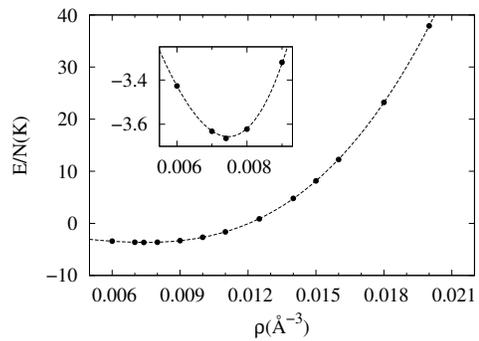}
\\  
        \caption{Energy per particle of liquid T$\downarrow$ (solid circles) as a
	function of the density $\rho$.  The solid line corresponds to the 
	fit to the DMC energies using 	Eq. (\protect\ref{eqestat}).   
	The error bars of the DMC energies are smaller than the size of the symbols.  }
     \label{fig:enliq}	
\end{figure}

In Fig. \ref{fig:enliq}, we plot the DMC results for the equation of state
of the liquid. We have tried different analytical forms to fit the DMC
data.  The best results have been obtained  by using a polynomial fit of the
form ($e \equiv E/N$)
\begin{equation}
e(\rho)  = e_0+B\left(\frac{\rho-\rho_0}{\rho_0}\right)^2+
C\left(\frac{\rho-\rho_0}{\rho_0}\right)^3 \ ,
\label{eqestat}
\end{equation}
$\rho_0$ and $e_0$ being the equilibrium density and the energy per
particle at this density, respectively. The equation of state
(\ref{eqestat}) is shown as a solid line on top of the DMC data in Fig.
\ref{fig:enliq}. The best set of parameters is: $e_0=-3.656(4)$ K,
$B=6.86(7)$ K ,  $C=4.70(5)$ K , and  $\rho_0=0.007466(7) $ \AA$^{-3}$, the
figures in parenthesis being the statistical uncertainties. 
It is worth noticing that the value obtained for the equilibrium density
expressed in units of $\sigma^{-3}$, $\rho_0=0.369~ \sigma^{-3}$ ($\sigma=
3.67$~\AA) is similar
to the one in liquid $^4$He, $\rho_0=0.365~ \sigma^{-3}$ ($\sigma=2.556$ 
\AA).  

\begin{figure}
\centering
        \includegraphics[width=0.8\linewidth]{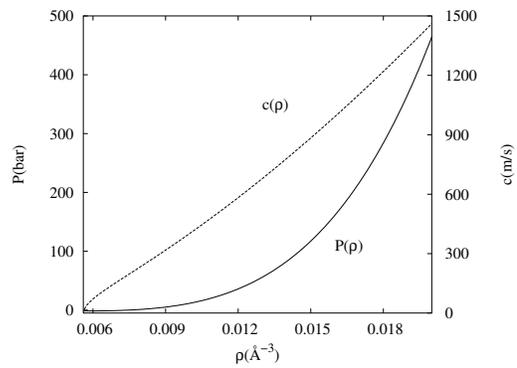}%
        \caption{Pressure and speed of sound of liquid T$\downarrow$ 
	as a function of the density. Left (right) scale corresponds to
	pressure (speed of sound).} 
	\label{fig:presliquid}
\end{figure}

Using the equation of state (\ref{eqestat}), we have obtained the pressure 
from its thermodynamic definition 
\begin{equation}
P(\rho)=\rho^2(\partial e/ \partial \rho) \ ;
\label{pressure}
\end{equation}
and from it, the corresponding speed of sound as a function of the density
\begin{equation} 
c^2(\rho)=\frac{1}{m} \left( \frac{\partial P}{\partial \rho} \right) \ .
\label{speed}
\end{equation} 
In Table \ref{tab:Liquid}, we report results for the pressure $P$ and the
speed of sound $c$ for some values of the density, where specific DMC
simulations have been carried out. The functions $P(\rho)$ and $c(\rho)$,
derived respectively from Eqs. (\ref{pressure}) and (\ref{speed}), are
shown in Fig. \ref{fig:presliquid}.

\begin{figure}
\centering
        \includegraphics[width=0.8\linewidth]{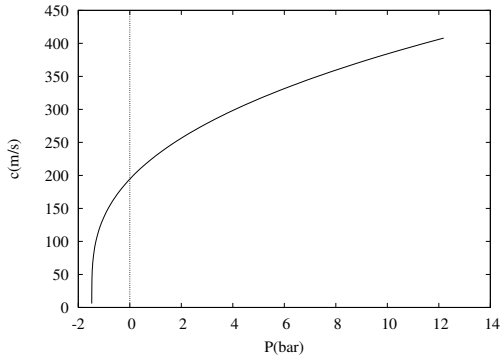}%
        \caption{Speed of sound of liquid T$\downarrow$ as a function of
	the pressure from the spinodal point up to freezing. The spinodal
	pressure $P_{\text s}$ corresponds to the point where	$c=0$.} 
	\label{fig:cofp}
\end{figure}

The spinodal density in T$\downarrow$, i.e., the density where the speed of
sound becomes zero,  is $\rho_0=0.0056$ \AA$^{-3}=0.277$  $\sigma^{-3}$, 
corresponding to a  pressure of $P_{\text s}= -1.48(2)$ bar. For
comparison, the spinodal density in liquid $^4$He is a bit lower,
$\rho_0=0.264$ $\sigma^{-3}$, the spinodal pressure being larger in absolute
value, $P_{\text s}= -9.30(15)$ bar. In Fig. \ref{fig:cofp}, we plot the
speed of sound $c$ (\ref{speed}) as a function of the pressure for pressures approximately up
to solidification. It can be seen that $c$ drops very fast when approaching
the spinodal point. Near the spinodal point it is expected that $c$ has the
form $\propto(P-P_{\text s})^{1/\nu}$, where $\nu$ is the critical exponent.

\begin{figure}
\centering
        \includegraphics[width=0.75\linewidth]{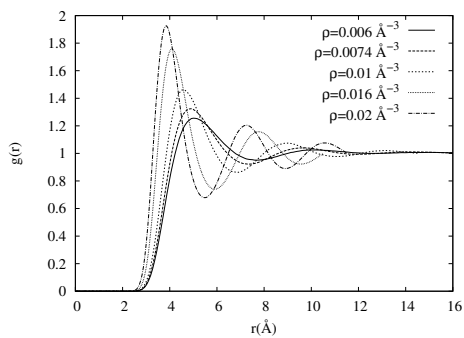}%
        \caption{Two-body radial distribution functions of the liquid phase. 
	From bottom to top in the height of the main
	peak, the results correspond to densities 0.006, 0.0074,  0.01,
	0.016, and 0.02 \AA$^{-3}$.  }
	\label{fig:grfluid}
\end{figure}

Apart from the ground-state energy, DMC simulations enables us to make
 exact estimations
of other relevant magnitudes such as the two-body radial distribution
function $g(r)$ and its Fourier transform, the static structure function
$S(k)$. The use of  pure estimators~\cite{pures} eliminates  the bias
coming from the trial wave function and allows us to arrive to exact
results for both functions. The evolution of $g(r)$ with density for 
liquid T$\downarrow$ is shown  in Fig. \ref{fig:grfluid}.  It is very
similar to the well-known results for liquid $^4$He. 
When $\rho$ increases, $g(r)$ gains structure, with the
main peak  shifting to shorter distances and increasing its strength.   

\begin{figure}
\centering
        \includegraphics[width=0.75\linewidth]{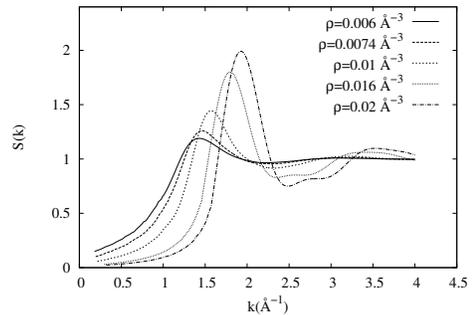}%
        \caption{Static structure function of the liquid phase. 
	From bottom to top in the height of the main
	peak, the results correspond to densities 0.006, 0.0074, 0.01,
	0.016, and 0.02 \AA$^{-3}$.}
	\label{fig:sofk}
\end{figure}

In Fig. \ref{fig:sofk}, results of $S(k)$ at the same densities as in
Fig. \ref{fig:grfluid} are reported.  The results again show the expected behavior: 
with the increase of $\rho$, 
the strength of the main peak increases and moves to higher momenta  
in a monotonic way. At low momenta, the slope of $S(k)$ decreases with the
density, following the limiting behavior $\lim_{k \rightarrow 0} S(k)= \hbar
k/(2mc)$ driven by the speed of sound $c$. The DMC data start at a 
finite $k$ value inversely proportional to the box size $L$.    

\begin{figure}
\centering
        \includegraphics[width=0.7\linewidth]{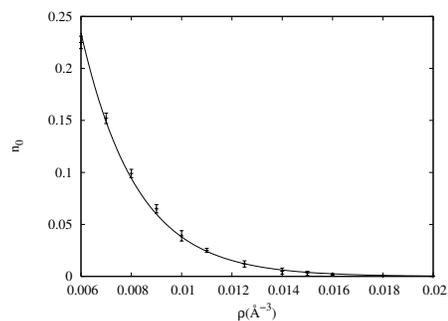}%
        \caption{Condensate fraction of spin-polarized T in the liquid phase. 
	The line corresponds to a fit to the DMC data using
	Eq.(\protect\ref{nofit}). The
	error bars are smaller than the size of the symbols.}
	 \label{fig:noliq}
\end{figure}

A characteristic signature of bulk superfluidity is the finite value of
the condensate fraction $n_0$, i.e. the fraction of particles occupying the
zero-momentum state. It has been extracted from the long-range behavior of
the one-body density matrix $\rho(r)$, by means of the asymptotic condition
$n_0=\lim_{r \rightarrow \infty} \rho(r)$. We have verified, by increasing
the number of particles of the simulation at different densities, that the
size dependence of $n_0$ is smaller than its statistical error. The results
obtained using the extrapolated estimator $n_0 \simeq 2 \langle n_0
\rangle_{\text DMC} - \langle n_0 \rangle_{\text VMC}$ are presented in 
Fig. \ref{fig:noliq}. The line on top of the data
corresponds to the exponential fit
\begin{equation}
n_0(\rho)= A \exp(-b \rho)
\label{nofit}
\end{equation}
with $A$=3.6(3) and $b$=455(11) \AA$^{-1}$, which describes well the DMC data.
At the equilibrium density, $n_0$=0.129(3). For comparison, $^4$He at the equilibrium has $n_0$=0.084(1), as obtained by the DMC method using a Jastrow wavefunctions for importance sampling in Ref. \onlinecite{boro}.

Previous theoretical estimates of the energy per particle in bulk tritium were obtained by
VMC~\cite{etters} or Brueckner-Bethe-Goldstone formalism.~\cite{tritium}
The T$\downarrow$-T$\downarrow$ interaction was modeled with the Morse
potential and the calculations were performed for densities up to
$0.015$ \AA$^{-3}$. Etters  {\it et al.}~\cite{etters}  obtained an
equilibrium energy per particle $E/N =-0.75(2)$ K for the equilibrium
density $\rho_0=0.0463$ \AA$^{-3}$. For the same density, Joudeh et {\it
al.}~\cite{tritium} obtained $-0.759$ K. Comparing this equilibrium density
with the present DMC results one notices that this variational estimate
lies below our estimated spinodal point. In order to compare our results with
previous ones, we have carried out simulations at a density
$\rho=0.005$ \AA$^{-3}$ using both the Morse potential and the JDW one.
With the Morse
potential and at the VMC level we obtain $E/N=-0.74(1)$ K, which is within the
errorbars the same result as that of Etters {\it et al.}($-0.75(2)$ K). Using the
Morse potential, the 
DMC calculation  lowers the energy to $-0.978(1)$ K.  At the same
density $0.005$ \AA$^{-3}$, we have also used  the JDW potential obtaining at
the DMC level a sizeable lower energy of $-2.900(2)$ K. 
The observed differences in the equation of
state imply also substantial changes in the estimated pressure and
compressibility. For example, for a density  $\rho=0.01$ \AA$^{-3}$ we find
a pressure of $11.6(1)$ bar, while in Ref. \onlinecite{etters} it is estimated
to be 17 bar. At higher densities, the difference between the pressure results grows,
amounting to approximately 80 bar at $0.015$ \AA$^{-3}$. 

Using the relation for the transition temperature of the ideal Bose-Einstein gas
\begin{equation}
T_c=3.31\left(\frac{\hbar^2}{mk_B} \right)\rho_0^{2/3} \ ,
\label{tc}
\end{equation} 
Etters et {\it al.}~\cite{etters} estimated the temperature of superfluid
transition of spin-polarized T to be $1.48$ K. With the same argument, our
results for the equilibrium density suggest a superfluid transition
temperature of $2.02$ K. This is of course only an approximate  estimation
since in  liquid $^4$He Eq. \ref{tc} gives 3.1 K instead of the right one of $2.17$ K.

\subsection{Solid phase}

We have performed calculations of the
spin-polarized T solid phase with three different lattices (bcc, fcc, hcp)
and using the Nosanow-Jastrow wave function (\ref{nosanow}) for importance sampling. 
The energy per
particle in bcc, fcc, and hcp lattices has been obtained for different
densities in the range from $0.008$ \AA$^{-3}$ to $0.024$ \AA$^{-3}$. 
Our DMC results show that the energies per particle for the three lattices
are statistically indistinguishable in all the studied density regime. 
 Still, it reaches the lowest values in the hcp
solid phase, so this lattice seems to be energetically preferred. As an
example, the results at two densities for bcc, fcc and hcp lattices
are respectively: for $\rho=0.011$ \AA$^{-3}$, near the melting density,
the energies per particle are $E/N = -1.96(5)$, $ -1.98(7)$, and
$-2.04(6)$ K; at a higher density $\rho=0.018$ \AA$^{-3}$, 
 $E/N = 15.8(2)$, $ 15.3(2)$, and $15.26(8)$ K. The same
behavior has been observed for all densities  and therefore we
decided to investigate solid T$\downarrow$ properties assuming its hcp
crystalline structure. It is important to notice that the bcc lattice 
proved to be energetically preferred in a recent study of  the gas-solid phase 
transition in H$\downarrow$ 
\cite{hydrogen}  as well as in a study of solid hydrogen at very high pressure
\cite{ceperley}.

\begin{table}
\begin{center}
\begin{ruledtabular}                 
\begin{tabular}{lrrrr}
   $\rho$ (\AA $^{-3}$) & ~~$E/N$ (K) 
& ~~$T/N$ (K) & ~~$P$ (bar) & $c$ (m/s)  \\ \hline
0.01      &   -2.56(6)      &  14.16(8)    & 4.7(7)      &  321(18)   \\
0.011     &   -2.04(6)      &  16.51(9)    &  11(1)      &  404(19)   \\
0.015     &    4.17(8)      &  27.89(11)    &  83(4)      & 782(28)    \\
0.018     &   15.26(8)      &  38.30(12)    &  218(8)     & 1113(36)  \\     
0.024     &   62.59(8)      &  62.54(13)    &  902(29)    & 1898(58)               
 \end{tabular}
  \end{ruledtabular}
  \end{center}         
 \caption{Results for solid T$\downarrow$ at different densities $\rho$: 
energy per particle ($E/N$), kinetic energy per particle ($T/N$), pressure 
($P$), and speed of sound ($c$). Figures in parenthesis are the 
statistical errors.}
\label{tab:HCPtritium}
\end{table}

\begin{figure}
\centering
        \includegraphics[width=0.75\linewidth,angle=0]{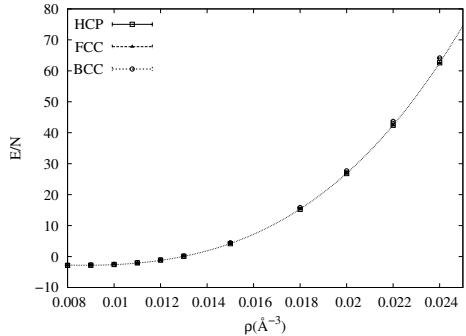}
        \caption{Energy per particle of solid T$\downarrow$ as a
	function of the density $\rho$ for the hcp (boxes), fcc(triangles)
	and bcc(circles) 
	lattices.  The solid line corresponds to the 
	fit to the DMC energies using 	Eq. (\protect\ref{eqestatsolid}).   
	The error bars of the DMC energies are smaller than the size of the symbols.  }
\label{fig:ensol}	
\end{figure}

In Fig. \ref{fig:ensol}, the DMC energies per particle of the solid phase
at different densities have been shown, for the three lattices. The line on
top of the data in the figure corresponds to the equation of state of the solid hcp
lattice. It has been obtained by fitting the DMC results with a polynomial
function of the form  
\begin{equation}
e(\rho) =  s_2\rho^{2} +  s_3\rho^{3} +  s_4\rho^{4}  \ ,
\label{eqestatsolid}
\end{equation}
with parameters $s_2=-10.47(11)\times 10^{4}$ K\AA$^{2}$, $s_3=7.13(15)\times
10^{6}$ K\AA$^{3}$, and $s_4=7.3(5)\times 10^{7}$ K\AA$^{4}$.

The total and kinetic energies per particle for several selected
densities are given in Table \ref{tab:HCPtritium}. Kinetic energy is 
determinated in the same way as in the liquid phase. The total energy is
negative for  $\rho \leq 0.012$ \AA$^{-3}$, while for greater
densities the kinetic energy exceeds the absolute value of the potential
one
causing the total energy to become positive. The potential energy is
negative for all the considered densities, with a single  exception
corresponding to the highest density $0.024$  \AA$^{-3}$. As it is reported in
Ref. \onlinecite{hydrogen}, the smaller mass of H$\downarrow$ atoms caused
the potential
energy to enter in regime of positive values for slightly smaller densities
($\rho \geq 0.02$ \AA$^{-3}$).

\begin{figure}
\centering
        \includegraphics[width=0.75\linewidth,angle=0]{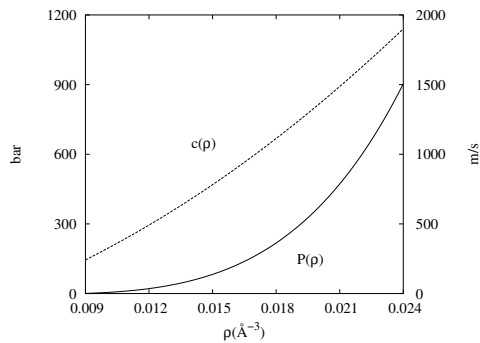}%
        \caption{Pressure and speed of sound of solid T$\downarrow$ 
	as a function of the density. Left (right) scale corresponds to
	pressure (speed of sound).} 
	\label{fig:pressolid}
\end{figure}

In
Fig.\ref{fig:pressolid}, we show the  pressure and speed of sound of solid T$\downarrow$
obtained from the equation of state (\ref{eqestatsolid}) using the
thermodynamic relations  (\ref{pressure}) and (\ref{speed}) as a
function of the density. Comparison with values of the same quantities in
solid bulk H$\downarrow$, reported in Ref. \onlinecite{hydrogen}, reveals smaller
pressure and speed of sound  for solid T$\downarrow$ at the same densities.

\begin{figure}
\centering
        \includegraphics[width=0.75\linewidth,angle=0]{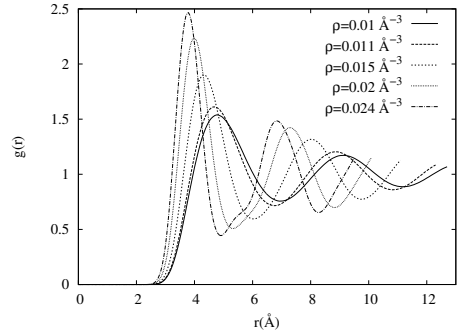}%
        \caption{Two-body radial distribution functions of the solid phase. 
	From bottom to top in the height of the main
	peak, the results correspond to densities 0.01, 0.011, 0.015, 0.02
	and 0.024 \AA$^{-3}$.  }
	\label{fig:grsolid}
\end{figure}

The spatial pattern of the solid structure is reflected in the two-body radial
distribution function $g(r)$. In  Fig. \ref{fig:grsolid}, we have plotted
$g(r)$ for same selected densities. The strength  of the main peaks are greater in
the solid than in the liquid phase, as can be seen by comparison of our
results at 
densities $0.01$ \AA$^{-3}$  and $0.02$ \AA$^{-3}$. Also, it is clear that
secondary peaks are larger in the solid phase. Just like in the liquid, 
when the density increases the height of the main peaks grows and moves
to shorter distances.

\subsection{Liquid-solid phase transition}

\begin{figure}
\centering
        \includegraphics[width=0.7\linewidth]{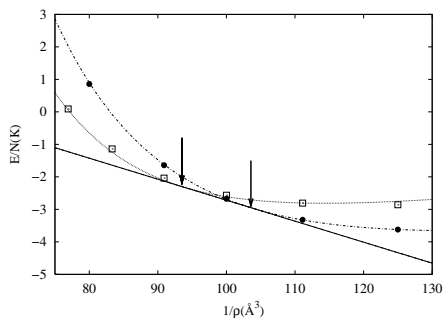}%
        \caption{Maxwell construction based on plotting the energy per
particle, $E/N$ as a function of $1/\rho$. The densities at which the first-order 
transition occurs are identified by finding the common tangent (solid line)
to both the
solid (dotted line) and liquid curve (dot-dashed line). }
\label{fig:Max}
\end{figure}

An important information that can be derived from the DMC simulations of
the liquid and solid phases is the location of the liquid-solid phase transition
point.
 As in a recent investigation of the gas-solid phase transition in
H$\downarrow$ \cite{hydrogen}, we have used  the double-tangent Maxwell
construction to determine the transition.  This well-known method is based on 
the search of a common tangent to
both the liquid and solid equations of state whose intersections give 
the freezing ($\rho_ {\rm f} $) and melting ($\rho_{\rm m}$) densities, 
as plotted in Fig. \ref{fig:Max}. 
Using the equation of state of the hcp crystalline structure, we have
obtained $\rho_ {\rm f} =0.00964$ \AA$^{-3} =0.477$ $\sigma^{-3}$ and  
$\rho_{\rm m} =0.01069$ \AA$^{-3}=0.528$ $\sigma^{-3}$, which corresponds to a
common pressure at the phase transition of $P=9(1)$ bar. In addition, in order to
estimate the influence of the lattice type on the transition pressure, we
have also calculated the transition densities and pressure for fcc and bcc
lattices. For a fcc lattice $P=9.5(1.0)$ bar, while for bcc we have obtained
$P=9.9(1.0)$ bar. The three transition pressures fall within the errorbars,
but since hcp is consistently lower than the others it leads us to conclude
that it may be slightly preferred, like it is preferred in $^4$He.  The
transition densities in $^4$He are $\rho_ {\rm f} =0.430\ \sigma^{-3}$ and
$\rho_ {\rm m}=0.468\ \sigma^{-3}$,~\cite{edwards} corresponding to a pressure of 25.3 bar. 
The comparison can also be made with with gas-solid transition in
spin-polarized H$\downarrow$~\cite{hydrogen}. In this case,  the phase
transition occurs at higher densities
$\rho_ {\rm f} =0.01328$ \AA$^{-3}$, $\rho_ {\rm m} =0.01379$ \AA$^{-3}$
and much higher pressure, $P=173(15)$ bar, than in spin-polarized T. This
effect can be explained as a consequence of its isotopic difference since
T$\downarrow$ atoms have approximately three times greater mass than
H$\downarrow$ atoms.

As in any first-order phase transition, the density is discontinuous in the
transition point.
 Another quantity which is also not continuous crossing the phase 
transition  is the kinetic energy per particle. Namely, near the freezing density 
the kinetic energy per particle of the liquid is 
$T/N=13.06(3)$ K, while at the same time, near the melting density but in the solid
phase the same energy is $T/N=16.51(3)$ K. Therefore, our results for bulk 
T$\downarrow$ show a discontinuity of the kinetic energy of  around 3.5
K; in bulk H$\downarrow$, the same difference has been shown to be  more than
twice times larger.\cite{hydrogen}  On the other hand, the  value of the 
condensate fraction in
the two spin-polarized systems at the corresponding transition densities
are similar, 0.03 for T$\downarrow$ and 0.04 for H$\downarrow$~\cite{hydrogen}.

\begin{figure}
\centering
        \includegraphics[width=0.75\linewidth,angle=0]{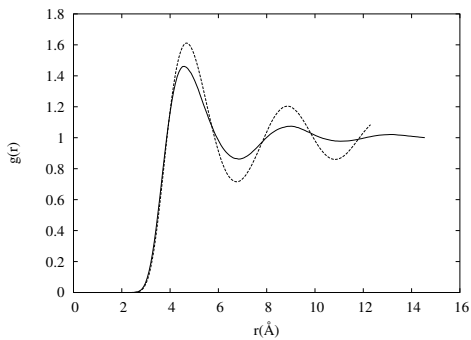}%
        \caption{Two-body radial distribution function at the liquid-solid
	phase transition. The solid line corresponds to the liquid at
	$\rho_{\rm f}$ and the dashed line to the solid at $\rho_{\rm m}$.}
\label{fig:grtrans}
\end{figure}

A measure of the mean displacement of T$\downarrow$ atoms
with respect to the lattice sites is obtained by computing the Lindemann's
ratio ($\gamma$). 
As usual, we calculate it by sampling
the expression  $\gamma=\sqrt{\langle (\bm{r}-\bm{r}_I)^2\rangle}
/a_{L} $, where $a_L$ is the lattice constant. The parameter $\gamma$
decreases monotonically with increasing density and the values we have
obtained show a slight dependence on the particular lattice chosen for the
simulation: it is  greatest for hcp and
smallest for bcc. Particularly, at the melting density of the hcp phase
Lindemann's ratio assumes a value $\gamma=0.26$, which is the same as in
$^4$He and similar to the value estimated for solid H$\downarrow$
($\gamma=0.25$)~\cite{hydrogen}.

\begin{figure}
\centering
        \includegraphics[width=0.75\linewidth,angle=0]{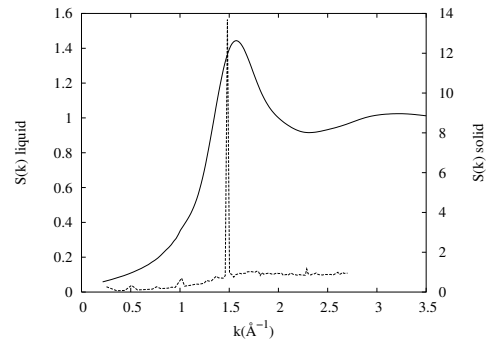}%
        \caption{Static structure factor at the liquid-solid
	phase transition. The results correspond to the liquid at
	$\rho_{\rm f}$ and to the solid at $\rho_{\rm m}$.}
\label{fig:sktrans}
\end{figure}

The discontinuity in the  liquid-solid transition is also revealed in the difference
between $g(r)$ of the liquid and solid phases (Fig. \ref{fig:grtrans}). Even
more dramatic difference between the two phases is demonstrated in Fig.
\ref{fig:sktrans} where  DMC results of $S(k)$ are shown at $\rho_{\rm f}$ and
$\rho_{\rm m}$ .  $S(k)$ in the solid phase is characterized with strong peaks
at reciprocal lattice sites whereas  this behavior is clearly not observed
in the  $S(k)$  of the  liquid phase. Finally, it is worth noticing that the
main peak of solid $S(k)$ is slightly weaker in  T$\downarrow$ than in
 H$\downarrow$ (reported in Ref. \onlinecite{hydrogen}). The main reason for this
lies in the fact that the liquid-solid transition in T$\downarrow$ emerges at
lower densities than gas-solid transition in H$\downarrow$.

\section{Conclusions}

The ground-state properties of spin-polarized tritium T$\downarrow$ have
been accurately determined using the DMC method in both the liquid and
solid phases. The obtained results are based on the precise knowledge of the 
T$\downarrow$-T$\downarrow$ interatomic potential, which combined with the
accuracy of the DMC method, allowed for a nearly exact determination of the
main properties of the system. All the previous results on liquid
T$\downarrow$ were based on a Morse potential  and approximate
computational methods.  Hence, our predictions for equilibrium density and
energy per particle differ substantially from those. We find that 
the equilibrium density is
$\rho_0= 0.00747(1)$ \AA$^{-3}$, which expressed in units of $\sigma^{-3}$ is very
similar to the one of liquid $^4$He ($\rho_0=0.369~ \sigma^{-3}$ for T$\downarrow$,
$\rho_0=0.365~ \sigma^{-3}$ in $^4$He ). Previous predictions of the equilibrium
density lie below our predicted spinodal density. Despite a similar
equilibrium density, T$\downarrow$ has approximately half the equilibrium energy per
particle of liquid $^4$He as a consequence of its smaller mass
and shallower potential. 

At a high enough density the system freezes. We have studied the energetic
and structural properties of the solid phase for three lattices.
Differences in energies between the phases are almost indistinguishable,
but hcp seems to be slightly preferred over the fcc and bcc ones. From the
DMC equations of state of the liquid and solid phases, we have 
localized the  liquid-solid transition point  of  T$\downarrow$ 
for the first time, to the best of our knowledge.
At zero temperature, the phase transition occurs at $P=9(1)$ bar.

\acknowledgments
J. B.  acknowledges support from DGI (Spain) Grant No.
FIS2008-04403 and Generalitat de Catalunya Grant No. 2005SGR-00779. I.B and L.V.M. acknowledge
support from MSES (Croatia) under Grant No. 177-1770508-0493.
We also acknowledge the support of the Central Computing Services at
the Johannes Kepler University in Linz, where part of the computations was
performed.  In addition, the resources of the Isabella
cluster at Zagreb University Computing Centre (Srce) and Croatian National
Grid Infrastructure (CRO NGI) were used.


\begin{thebibliography}{32}

\bibitem{hydrogen} L. Vranje\v{s} Marki\'c, J. Boronat and J. Casulleras, 
Phys. Rev. B \textbf{75}, 064506 (2007).

\bibitem{StwaleyNosanow} W. C. Stwaley and L. H. Nosanow, Phys. Rev. Lett. 
\textbf{36}, 910 (1976).

\bibitem{fried} D. G. Fried \textit{et. al}, Phys. Rev. Lett. \textbf{81}, 3811 (1998).


\bibitem{becalkali}M. H. Anderson, J. R. Ensher, M. R. Matthews, C. E.
Wieman, and E. A. Cornell, Science \textbf{269}, 
198 (1995); K. B. Davis, M. O. Mewes, M. R. Andrews, N. J. van Druten, D.
S. Durfee, D. M. Kurn, and W. Ketterle , Phys. Rev. Lett. \textbf{75}, 
3969 (1995); C. C. Bradley, C. A. Sackett, J. J. Tollett, and R. G. Hulet,
Phys. Rev. Lett. \textbf{75}, 1687 (1995).


\bibitem{kolos} W. Kolos and L. Wolniewicz, J. Chem. Phys. \textbf{43},
2429 (1965); Chem. Phys. Lett. \textbf{24}, 457 (1974).

\bibitem{jamieson} M. J. Jamieson, A. Dalgarno, and
L. Wolniewicz, Phys. Rev. A \textbf{61}, 042705 (2000).

\bibitem{panoff}R. M. Panoff and J. W. Clark, Phys. Rev. B 
\textbf{36} 5527 (1987)

\bibitem{flynn} M. F. Flynn, J. W. Clark, E. Krotscheck, R. A. Smith, 
and R. M. Panoff, Phys. Rev. B \textbf{32}, 2945 (1985).

\bibitem{skjetne}
B. Skjetne and E. \O stgaard, J. Phys.: Condens. Matter \textbf{11} 8017 (1999).


\bibitem{miller}M. D. Miller, L. H. Nosanow, Phys. Rev. B \textbf{15}, 4376 (1977).

\bibitem{etters} R. D. Etters, J. V. Dugan, Jr., and R. W. Palmer, J. Chem. 
Phys. \textbf{62}, 313 (1975).

\bibitem{tritium}B. R. Joudeh, M. K. Al-Sugheir, H. B. Ghassib, 
Physica B \textbf{388}, 237 (2007).

\bibitem{blume} D. Blume, B. D. Esry, Chris H. Greene, N. N. Klausen,
 and G. J. Hanna, Phys. Rev. Lett. \textbf{89}, 163402 (2002).
 
 \bibitem{yan}Zong-Chao Yan, James F. Babb, A. Dalgarno, and G. W. F. Drake, 
Phys. Rev A \textbf{54}, 2824(1996).

\bibitem{kolos2}W. Kolos and J. Rychlewski, J. Mol. Spectrosc.
\textbf{143}, 237 (1990).
 
\bibitem{boro}J. Boronat and J. Casulleras, Phys. Rev. B \textbf{49} 8920 (1994).

\bibitem{overpressure} L. Vranje\v{s}, J.Boronat, J. Casulleras,
 and C. Cazorla, Phys. Rev. Lett. \textbf{95} 145302 (2005).

\bibitem{pures} J. Casulleras and J. Boronat, Phys. Rev. B \textbf{52}, 3654 (1995).

\bibitem{ceperley} C. Pierleoni, D. M. Ceperley, and M. Holzmann, Phys. Rev. Lett. 
\textbf{93}, 146402 (2004).

\bibitem{edwards}
 D. O. Edwards and R. C. Pandorff, Phys. Rev. A 140, 816
(1965).


\end{thebibliography}
\end{document}